# Revisiting the electron affinity of selenium

Rui Zhang, Wenru Jie, Jiayi Chen, Qihan Liu, and Chuangang Ning *

*Department of Physics, State Key Laboratory of Low Dimensional Quantum Physics,
Frontier Science Center for Quantum Information, Tsinghua University, Beijing 100084, China*

The electron affinity (EA) of atomic selenium, previously established as 16,297.276(9) cm$^{-1}$ based on the laser photodetachment microscopy (LPM) measurements in 2012, exhibited a significant deviation from other earlier experimental values, yet it remained the accepted reference standard for over a decade. In this letter, we re-examined the EA of Se using the slow-electron velocity-map imaging method and revealed a substantial deviation in the LPM result. Measurements for the different isotopes of Se and the energy-level splitting of the neutral Se atom's $^3P_2$–$^3P_1$ further verified the accuracy and robustness of our SEVI method. Based on these experimental evidences, we recommended a revised EA(Se) value of 16,297.78(4) cm$^{-1}$, which is in excellent agreement with the previous laser photodetachment threshold (LPT) experimental results.

***Introduction.*** Electron affinity (EA), defined as the energy difference between a neutral species and its corresponding anion in their respective ground states, is a fundamental parameter for understanding the behavior of atoms and molecules in diverse chemical and physical phenomena [1-3]. The precise measurement of EAs for elements in the Periodic Table remains an active research frontier in experimental physics and chemistry [4-8]. Notably, the most accurate EA measurement has been achieved to date is for the main group element oxygen (O): the laser photodetachment threshold (LPT) method yielded an EA value of 1.461112972(87) eV, or 11,784.6709 (8) cm$^{-1}$ [9].

Selenium (Se, atomic number Z = 34), a heavier group congener of oxygen, has an experimentally controversial EA value. The electron affinity of selenium was first measured as 16297(2) cm$^{-1}$ using the LPT method by H. Hotop *et al.* [10] in 1973. They also simultaneously determined the spin-orbit splitting between $^2P_{3/2}$ and $^2P_{1/2}$ states of Se$^-$ anion to be 2279(2) cm$^{-1}$. Subsequent measurements by N. B. Mansour *et al.* in 1988 [11] achieved tenfold improvement in precision, reporting EA(Se) = 16297.8(2) cm$^{-1}$ through tunable dye laser photodetachment spectroscopy combined with a Penning ion trap. Another independent LPT measurement by J. Thøgersen *et al.* [12] confirmed this EA value with a comparable precision [16,297.7(4) cm$^{-1}$], while refining the spin-orbital splitting $^2P_{3/2} - ^2P_{1/2}$ to 2278.2(2) cm$^{-1}$ in both single-photon and multiphoton absorption regimes. A notable discrepancy emerged when M. Vandevraye *et al.* [13] reported a significantly higher-precision EA(Se) value of 16,297.276(9) cm$^{-1}$ employing the laser photodetachment microscope (LPM) method, which deviates significantly from the previous LPT results. However, M. Vandevraye *et al.* claimed that investigating the reasons for a possible overestimation of detachment thresholds method was beyond the scope of their study, and they maintained their conclusion.

Despite the ongoing debate surrounding the electron affinity of selenium, the LPM measurement result has still remained the primary reference standard for over a decade. This persistent controversy of EA(Se) has motivated our re-examination of EA(Se) using slow-electron velocity-map imaging (SEVI) spectroscopy [14,15]. The SEVI method has demonstrated its robustness in many EA measurements thanks to its high energy resolution and ability to resolve multiple photodetachment channels. The SEVI technique has been successfully employed to accurately determine the EAs of the main group elements such as O, S, As, Sb, and Pb [8,16,17], achieving accuracies typical better than 0.1 cm$^{-1}$. In addition, this method has also been successfully applied to determine EAs across numerous transitional elements, lanthanides, and actinides, with typical accuracies of 0.01–0.1 meV [7,8,18-20]. Given these proven capabilities, we present a reinvestigation of selenium's electronic structure and its anion, trying to solve the discrepancy using the SEVI spectroscopy.

***Experimental Methods.*** The experiment was conducted on our Cryo-SEVI apparatus, as detailed elsewhere [21-23]. In this work, Se$^-$ anions were generated via laser ablation of a molten selenium target with helium carrier gas delivered through a pulsed valve under a backing pressure around 3 × 10$^5$ Pa. The produced anions were then accumulated in a radio frequency (RF) octupole ion trap [24], assisted by a RF hexapole guide. The trapped anions lost their kinetic energies through continuous collisions with the buffer gas helium for ~ 45ms. Subsequently, the anions were extracted from the ion trap and mass-selected via a time-of-flight spectrometer [25]. The selected Se$^-$ anions were then photodetached by a linearly polarized dye laser (Spectra-Physics) in the photodetachment region, with the wavelength monitored by a wavelength meter (HighFinesse WS6-600) with an accuracy of 0.02 cm$^{-1}$. The detached photoelectrons were analyzed via a velocity-map imaging (VMI) spectrometer [26,27], and the imaging voltages was set as –75V for the EA measurement. The distribution of outgoing photoelectrons can be reconstructed from the projected image using the maximum entropy velocity Legendre reconstruction (MEVELER) method [28] because the photoelectron distribution has a cylindric symmetry. The kinetic energy of the photoelectrons was given by $E_k = \alpha r^2$, where $r$ is the weighted center radius of reconstructed three-dimensional spherical shell and $\alpha$ is the calibration coefficient. The center position of the peak ($r$) was determined via Gaussian function fitting. The binding energy (BE) of each transition was calculated using Einstein's equation, $BE = h\nu - E_k$, where $h\nu$ is the photon energy of the detachment laser. The SEVI setup run at a repetition rate of 20 Hz. Each photoelectron spectrum was acquired by accumulating 50,000 laser shots typically.

*Contact author: ningcg@tsinghua.edu.cn

*High-resolution photoelectron spectra of Se$^-$.* Figure 1 shows the photoelectron spectra of Se$^-$ anions acquired at two different photodetachment wavelengths using a tunable dye laser. The spectra exhibit no observable contamination from hydride species SeH$^-$, despite Se has serval natural isotopes ($m$ = 74, 76, 77, 78, 80, 82). This spectral purity is attributable to the significant EA difference between SeH [2.21(3) eV] and Se [2.02067(2) eV], enabling complete spectral discrimination [29]. Figure 1 (b) presents the corresponding schematic energy-level diagram of anionic and neutral Se states. Three transitions from anionic Se$^-$ to neutral Se (labeled A–C) were observed in the photoelectron spectra. Peak A is the photodetachment transition from the anionic ground state Se$^-$ ($^2P_{3/2}$) to the neutral ground state Se ($^3P_2$). Its binding energy was preliminarily determined to be 16,297.78(64) cm$^{-1}$, directly representing the electron affinity of atomic selenium. More precise measurements of this EA value will be detailed in the next subsection. Peak B is the transition from the anionic ground state Se$^-$ ($^2P_{3/2}$) to the neutral first excited state Se ($^3P_1$), and its binding energy was determined as 18,287.26(83) cm$^{-1}$. Peak C is the transition from the first excited state of Se$^-$ ($^2P_{1/2}$) to the first excited state of Se ($^3P_1$), with a binding energy determined as 16,011(1) cm$^{-1}$. The weak intensity of peak C reflects the low population of the anionic excited state $^2P_{1/2}$. It remains detectable thanks to the high energy resolution and high sensitivity of our SEVI spectrometer. By subtracting the binding energies of transitions B and C, the spin-orbital splitting of $^2P_{1/2}$–$^2P_{3/2}$ of negative ion was determined to be 2276 (2) cm$^{-1}$, which is consistent with the previously reported high-precision value of 2278.2(2) cm$^{-1}$ [12]. The peak widths of peaks A and B, fitted with Gaussian functions, demonstrates our instrument's resolution: the full width at half maximum (FWHM) of peak A is 0.64 cm$^{-1}$ with kinetic energy $E_k$ = 1.51 cm$^{-1}$. Similarly, the FWHM of peak B is 0.83 cm$^{-1}$ with $E_k$ = 3.11 cm$^{-1}$. The energy resolution of our SEVI apparatus is comparable with the values reported by Wang group (FWHM = 1.2 cm$^{-1}$ at $E_k$ = 5.2 cm$^{-1}$) [27] and by Neumark group (FWHM = 1.1 cm$^{-1}$ at $E_k$ = 1.8 cm$^{-1}$) [30].

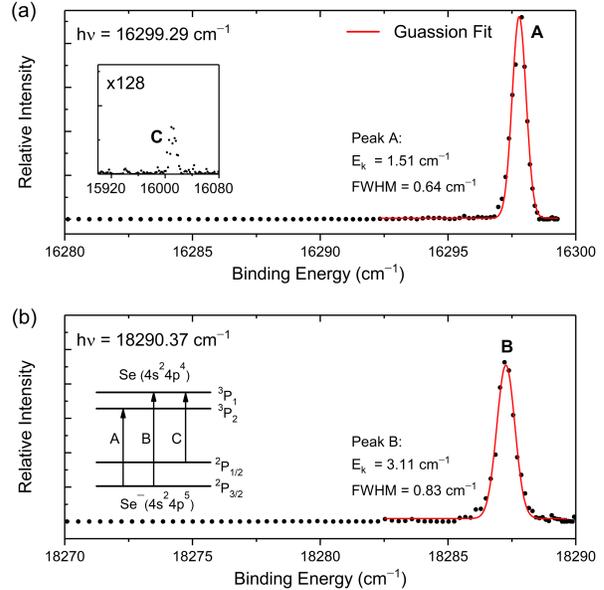

FIG. 1. Photoelectron spectra of Se$^-$ obtained at photon energies (a) 16 299.29 cm$^{-1}$ and (b) 18 290.37 cm$^{-1}$. The inset in (a) shows the magnified view of peak C, and inset in (b) shows the assignment of peaks A–C. The kinetic energies ($E_k$) and FWHMs for peaks A and B are also labeled.

*Precise measurement of the electron affinity of Se.* To accurately determine the electron affinity of Se, the photon energy $h\nu$ of the detachment laser was carefully tuned slightly above the photodetachment threshold of transition A. Sequential spectra spanning from 16,299 cm$^{-1}$ to 16,305 cm$^{-1}$ were acquired at 1 cm$^{-1}$ intervals. Since the kinetic energy of photoelectrons is proportional to the squared radius ($r^2$) of photoelectron sphere, a linear regression analysis on the $h\nu$ versus $r^2$ was implemented in Fig. 2 based on the Einstein's photoelectric equation $h\nu = BE + \alpha r^2$. The binding energy of the transition Se ($^3P_2$) ← Se$^-$ ($^2P_{3/2}$) was given by the vertical-axis intercept. This analysis yielded a refined EA(Se) value of 16,297.78 ± 0.04 cm$^{-1}$, equivalent to 2.020 667(5) eV using the 2018 CODATA conversion (1 eV = 8065.543 937... cm$^{-1}$) [31]. The total uncertainty 0.04 cm$^{-1}$ includes a systematic uncertainty 0.02 cm$^{-1}$ from the wavelength meter. To evaluate the isotope effects on the EA measurement for Se atom, we measured the binding energies of transition A for five primary isotopes of selenium: $^{76}$Se (abundance 9%), $^{77}$Se (8%), $^{78}$Se (24%), $^{80}$Se (50%) and $^{82}$Se (9%). Our measurement suggested that the isotope shifts remain below 0.03 cm$^{-1}$, within experimental uncertainty. Additionally, we also changed the imaging voltage from −75V to −150V. No observable difference was detected at two different imaging voltages. This further excludes so-called "quantum offset" effects [32] in the SEVI experiments, which incorrectly claimed that SEVI results have a field-dependent systematic deviation as large as 0.2 cm$^{-1}$. Table I listed the measured EA(Se) values for various selenium isotopes and the corresponding imaging voltages.

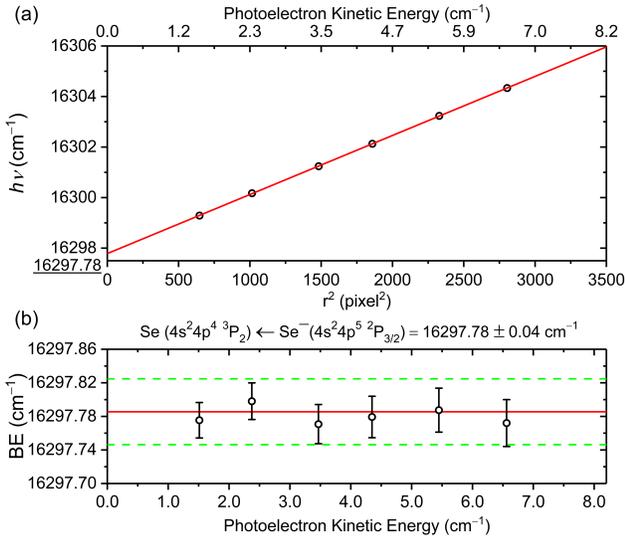

FIG. 2. (a) Photon energy ($hv$) versus squared radius ($r^2$) of photoelectron spherical shells for the transition A. The red solid line represents a linear least-squares fit, where the intercept, 16 297.78 cm$^{-1}$, is the measured EA value of selenium. (b) The binding energy (BE) of transition A as a function of the photoelectron kinetic energy. The green dashed lines represent ± 0.04 cm$^{-1}$ uncertainty.

TABLE I. Measured EA(Se) values for various isotopes of Se with two imaging voltages.

| Isotope | Imaging Voltage/V | EA(Se)/cm$^{-1}$ |
|---|---|---|
| $^{76}$Se | 150 | 16297.77(6) |
| $^{76}$Se | 75 | 16297.78(4) |
| $^{77}$Se | 75 | 16297.76(4) |
| $^{78}$Se | 75 | 16297.78(4) |
| $^{80}$Se | 75 | 16297.79(4) |
| $^{82}$Se | 75 | 16297.79(4) |

*Verification via an alternative transition.* To verify the reliability of our EA(Se) measurement, we also accurately measured the binding energy of transition B $^2P_{3/2}$ (Se$^-$)– $^3P_1$(Se) using the same method. Since transitions A and B are both from the same anionic ground state, the energy difference between these two transitions equals the spin-orbital splitting of the final states Se ($^3P_2$) and Se ($^3P_1$), which is well known to be 1989.497 cm$^{-1}$ according to the NIST atomic data [33,34]. As shown in Fig. 3, the squared radius $r^2$ was plotted versus the photon energy $hv$ ranging from 18,289 cm$^{-1}$ to 18,294 cm$^{-1}$. The binding energy of the transition Se ($^3P_1$) ← Se$^-$ ($^2P_{3/2}$) was determined to be 18,287.28(5) cm$^{-1}$. The $^3P_2$–$^3P_1$ energy-level splitting of Se was obtained by subtracting the binding energies of transitions A from that of B, yielding a value of 1989.50(7) cm$^{-1}$. This splitting value is in excellent agreement with the NIST reference result 1989.497 cm$^{-1}$, which further confirms the reliability of our experimental results of EA(Se). In other words, this alternative channel also yielded EA(Se) = 16,297.78(5) cm$^{-1}$ by subtracting the reference splitting 1989.497 cm$^{-1}$ from the binding energy 18,287.28(5) cm$^{-1}$, which is in excellent agreement with the value 16,297.78(4) cm$^{-1}$ obtained via the transition Se ($^3P_2$) ← Se$^-$ ($^2P_{3/2}$). All measured binding energies of transition A-C are recorded in Table II.

Figure 4 compares experimental EA values of Se with the previously reported results. For clarity, the early LPT result, EA(Se) = 16,297(2) cm$^{-1}$ by H. Hotop *et al.* in 1973, is omitted due to their substantially large error bar making the other small error bar invisible when plotted in the same scale [10]. Our SEVI measurement result is consistent with the values of J. Thøgersen *et al.* [12] and N. B. Mansour *et al.* [11] using the LPT method. The accuracy of our SEVI result represents an improvement of an order of magnitude over previous LPT measurements. A striking discrepancy emerges when comparing with the previously accepted benchmark LPM EA(Se) = 16,297.276(9) cm$^{-1}$ reported by M. Vandevrage *et al.* [13]. Their reported value deviates by 0.50 cm$^{-1}$ from the consensus—50 times greater than their claimed uncertainty.

Based on the above evidences, we can safely conclude that there is a significant deviation in the previous LPM result. Actually, M. Vandevrage *et al.* [13] pointed out that their experimental methodology employed for EA measurements exhibited systematic energy shifts correlated with the electric field in the photodetachment region. This estimated electric-field uncertainty reached approximately 2.5% in their LPM measurements of EA(Se), while their earlier EA measurements on phosphorus (P) demonstrated a markedly lower field-induced error of −0.3% [35]. M. Vandevrage *et al.* [13] acknowledged that this discrepancy between these two systematic errors observed in LPM measurements remained a mystery.

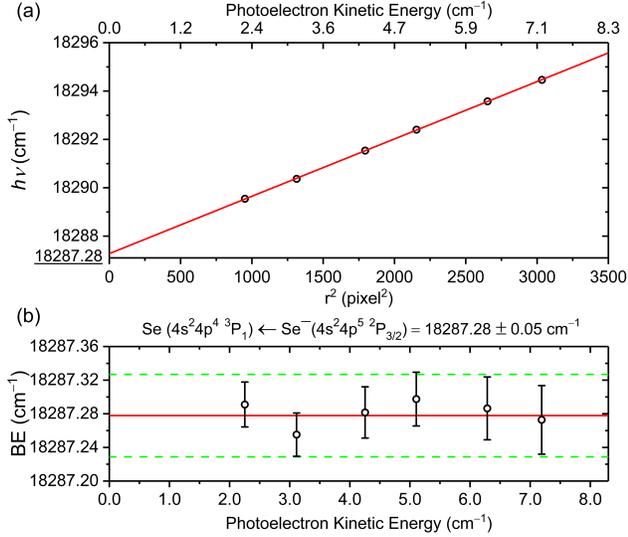

FIG. 3. (a) Photon energy ($h\nu$) vs squared radius ($r^2$) of photoelectron spherical shells for the transition B. The binding energy of transition B is determined to be 18 287.28 cm$^{-1}$ via a linear least-squares fit. (b) The binding energy (BE) of transition B as a function of the photoelectron kinetic energy. The green dashed lines represent ± 0.05 cm$^{-1}$ uncertainty.

TABLE II. Binding energies of observed transitions A–C.

| Peak | Transition (Se$^-$ → Se) [a] | Binding Energy/cm$^{-1}$ |
|---|---|---|
| A | $^2P_{3/2} \to \,^3P_2$ | 16297.78(4) |
| B | $^2P_{3/2} \to \,^3P_1$ | 18287.28(5) |
| C | $^2P_{1/2} \to \,^3P_1$ | 16011(1) |

[a] The electronic configuration of Se$^-$ is $4s^24p^5$, while that of Se is $4s^24p^4$.

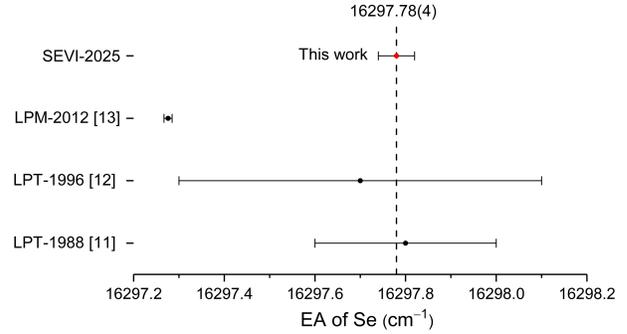

FIG. 4. The comparison of experimental EA(Se) values with error bars reported by different groups. The black dash line represents a recommendation EA(Se) value weighted by the SEVI and LPT measurements.

*Conclusion.* In summary, we have revisited the electron affinity of selenium with the SEVI method and pointed out a significant deviation of the previously reported LPM result of EA(Se). Based on our new SEVI results and earlier LPT measurements, we revised the new recommended value of EA(Se) as 16,297.78(4) cm$^{-1}$, resolving the longstanding discrepancies among prior EA(Se) measurements.

*Acknowledgments.* This work was supported by the National Natural Science Foundation of China (NSFC) (Grant Nos. 12374244 and 12341401).